\documentclass[10pt,aps,twocolumn,prl,noshowkeys,superscriptaddress,floatfix]{revtex4-1}

\usepackage{amsmath,amsfonts,amsthm,amssymb,mathrsfs}
\usepackage[dvips]{graphics,graphicx}
\usepackage{nicefrac,bm}
\usepackage[usenames,dvipsnames]{color}
\definecolor{darkblue}{RGB}{0,0,196}
\definecolor{darkgreen}{RGB}{0,120,0}
\usepackage[colorlinks=true,linktocpage=true,linkcolor=darkblue,citecolor=red,urlcolor=darkblue]{hyperref}
\usepackage{cancel}
\usepackage{multirow}
\usepackage{longtable}
\usepackage{color}
\usepackage[normalem]{ulem}
\usepackage{slashed}
\usepackage{hyperref}
\usepackage{bigints}
\usepackage{yfonts}
\usepackage{xparse}
\usepackage{physics}
\usepackage{verbatim}
\usepackage{minibox}
\usepackage{comment}
\usepackage{appendix}
\usepackage{nicefrac}

\begin{document}

\preprint{}

\title{Charge-selective topological locking and ultraslow coarsening in magnetoelastic media}

\author{Tirthankar Banerjee}
\email{tirthankar@niser.ac.in}
\affiliation{School of Physical Sciences, National Institute of Science Education and Research, An OCC of Homi Bhabha National Institute, Jatni-752050, India}

\author{Amaresh Jaiswal}
\email{a.jaiswal@niser.ac.in}
\affiliation{School of Physical Sciences, National Institute of Science Education and Research, An OCC of Homi Bhabha National Institute, Jatni-752050, India}

\date{\today}


\begin{abstract}
We demonstrate that the symmetry of magnetoelastic coupling governs the topological organization of composite defects and can qualitatively alter phase-ordering kinetics. In a two-dimensional theory of an in-plane magnetic vector coupled to elastic strain, we find that integer magnetic vortices bind pairs of elementary elastic half-defects. Isotropic magnetostriction produces a core-induced elastic winding independent of the magnetic charge, giving defect binding without charge locking and leaving the magnetic sector with essentially XY-like coarsening. In contrast, quadrupolar magnetoelastic coupling locks magnetic and elastic defect charges: continuous rotational symmetry favors $q_\phi=q_\theta$, whereas crystalline anisotropy can reverse the locking to $q_\phi=-q_\theta$. We reveal that both locking sectors display anomalously slow vortex annihilation: defect density shows a power-law decay $t^{-2/5}$, while the average defect-velocity  scales as $t^{-4/5}$. These results are consistent with a dynamic exponent $z\simeq 5$. We show that this coarsening slowdown is associated with large anisotropic elastic friction. An adiabatic elimination of the elastic field reveals that isotropic magnetostriction merely renormalizes local magnetic interactions, whereas orientational coupling induces effective nonlocal interactions between magnetic spins. Our experimentally testable results thus establish magnetoelastic coupling symmetry as a means of controlling defect topology and ordering kinetics.
\end{abstract}


\maketitle


%

\medskip
{\it Introduction:}
Topological defects are known to play a central role in the theory of phase ordering kinetics~\cite{Bray01061994, Mazenko1997}. In two-dimensional magnets with continuous orientational symmetry, this ordering is governed by the motion and annihilation of magnetic vortices, which sets the characteristic growth law of magnetic domains~\cite{Mazenko1997, PhysRevE.47.1525, PhysRevA.42.5865}. This familiar coarsening picture can be qualitatively altered when spin orientation is coupled to additional structural degrees of freedom, e.g., in magnetoelastic materials~\cite{Becker1939Ferromagnetismus, RevModPhys.21.541, Callen1963, Callen1965, doi:10.1126/science.aao0195}. Some interesting effects include deformation-induced enhancement of magnetic orientational order~\cite{PhysRevB.110.024312}, strong dynamical coupling between magnetic and shear degrees of freedom~\cite{Zhou2022}, and stabilization of the long-range magnetic ordering by magnetoelastic interactions~\cite{Fridman2002}. Experiments have shown that strain can be used to excite magnetic-vortex dynamics~\cite{PhysRevLett.133.146701}, generate topological textures non-thermally~\cite{Jani2024}, drive defect-mediated reorientation~\cite{PhysRevLett.134.136704}, and control magnetic anisotropy at the nanoscale~\cite{Kang2025, 10.1021/acsnano.4c16603}. Strong magnetoelastic effects have also been demonstrated across a growing range of magnetic materials~\cite{TANG2025100878, Cai2025, Ma_2025, Marques2025}, with direct relevance in manipulation of magnetic domains for information storage~\cite{Kriegner2016, Reimers2023}. Magnetism can also be used as a tool to control thin film coarsening~\cite{PhysRevMaterials.4.023404}. It is therefore timely to search for symmetry-governed topological principles that determine the evolution of coupled magnetic and elastic degrees of freedom. 

In this Letter, we consider a minimal two-dimensional continuum magnetoelastic  theory in which an in-plane magnetic order parameter ${\bf N}({\bf r},t)$ is coupled to the deformations ${\bf u}({\bf r},t)$ of an elastic medium. The magnetic orientation profile supports vortices and anti-vortices, while the principal axes of the strain define a second orientational field supporting half-integer defects.  Following a temperature quench, this coupled system develops complex orientational patterns containing composite topological defects, whose relative motion and  annihilation govern the growth of order.  

\begin{figure}
    \centering
    \includegraphics[width=0.98\columnwidth]{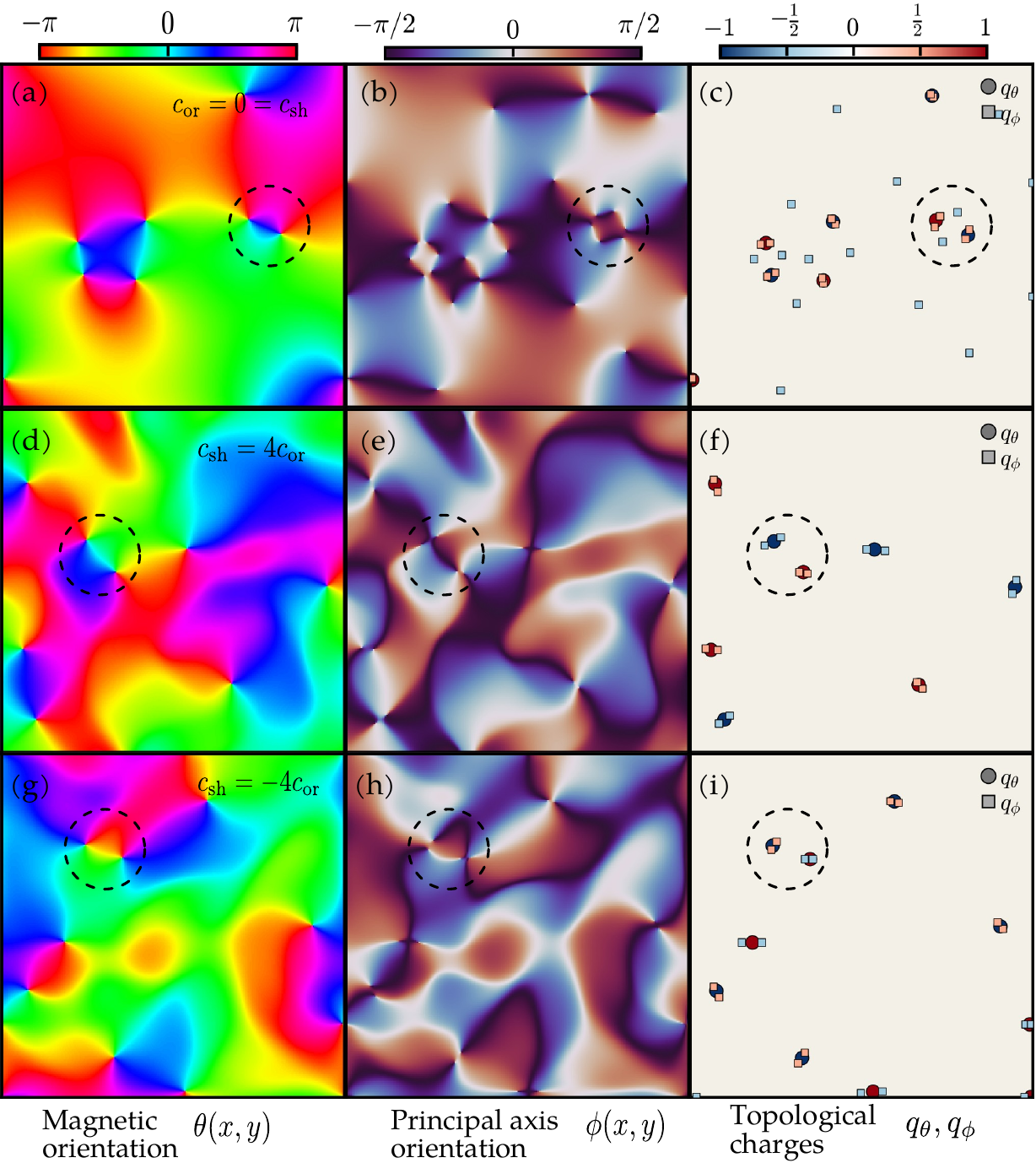}
    \caption{Representative snapshots of the magnetic orientation $\theta$ , elastic principal-axis orientation $\phi$ , and the corresponding topological defects at time $t=200$, for fixed isotropic magnetostriction $c_{\rm iso}=1$. Circles and squares denote magnetic ($q_\theta=\pm1$) and elastic ($q_\phi=\pm1/2$) defects. Colors indicate topological charges. Dashed circles highlight representative defect-annihilation clusters~\cite{Supp_Mov}.
(a)--(c) $c_{\rm or}=0=c_{\rm sh}$: Both magnetic vortex and anti-vortex bind to $+1/2$ elastic defects.
(d)--(f) $c_{\rm sh}=2=4 c_{\rm or}$: magnetic and elastic defects lock with the same sign of the respective topological charges.
(g)--(i) $c_{\rm sh} =-2= -4 c_{\rm or}$: charge-locking between magnetic and elastic defects with opposite signs.
Parameters: $L_x=64=L_y, dx=0.25, dt=0.01,\mu=1=2\lambda, k_1=0.5=k_2, a=-1=-b,v=1=\Gamma_N=\Gamma_u$.}
\label{fig:defects_all}
\vspace{-0.2cm}
\end{figure}

We show that the structure of the magnetoelastic coupling in the free energy determines how the magnetic and elastic defects communicate with each other. An isotropic magnetostrictive coupling responds primarily to the suppression of magnetic order near a vortex core and, therefore, produces an elastic distortion that is insensitive to the sign of the magnetic winding [Fig.~\ref{fig:defects_all}(a)]. By contrast, a quadrupolar coupling is sensitive to the relative orientation of magnetization and strain and can lock the local topological charges carried by the two fields [Fig.~\ref{fig:defects_all}(b,c)]. The nature of magnetoelastic couplings therefore realizes two qualitatively different forms of defect association: binding {\em with} or {\em without} charge-selective topological locking.

\begin{figure}
    \centering
    \includegraphics[width=0.9\columnwidth]{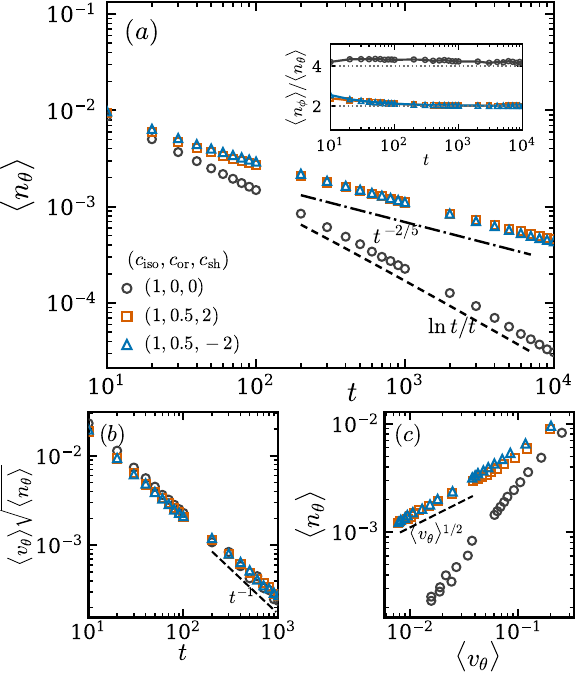}
    \caption{(a) Magnetic defect density $\langle n_\theta \rangle = \langle N_\theta/L_xL_y\rangle$ {\em vs.} time $t$ for different choices of coupling constants. Data averaged over $40$ realizations. Parameters: $L_x =512= L_y, dx=1, dt=0.01$. (b) $\langle v_\theta\rangle \sqrt{\langle n_\theta\rangle}$ {\em vs.} time $t$ for different values of coupling constants. (c) Parametric plot of $\langle n_\theta \rangle$ vs $\langle v_\theta\rangle$ shows that in the charge-selective locking regime (small $\langle v_\theta\rangle$), $\langle v_\theta\rangle \sim  \langle n_\theta\rangle^2$. Data in (b,c) averaged over $128$ realizations. Symbols have the same meaning as in panel (a). Parameters (b,c): $L_x = 256 = L_y, dx=0.5, dt=0.01$. All other parameters in (a) and (b) same as in Fig.~\ref{fig:defects_all}.}
    \label{fig:scaling}
    \vspace{-0.2cm}
\end{figure}

\begin{figure}
    \centering
     \includegraphics[width=0.99\columnwidth]{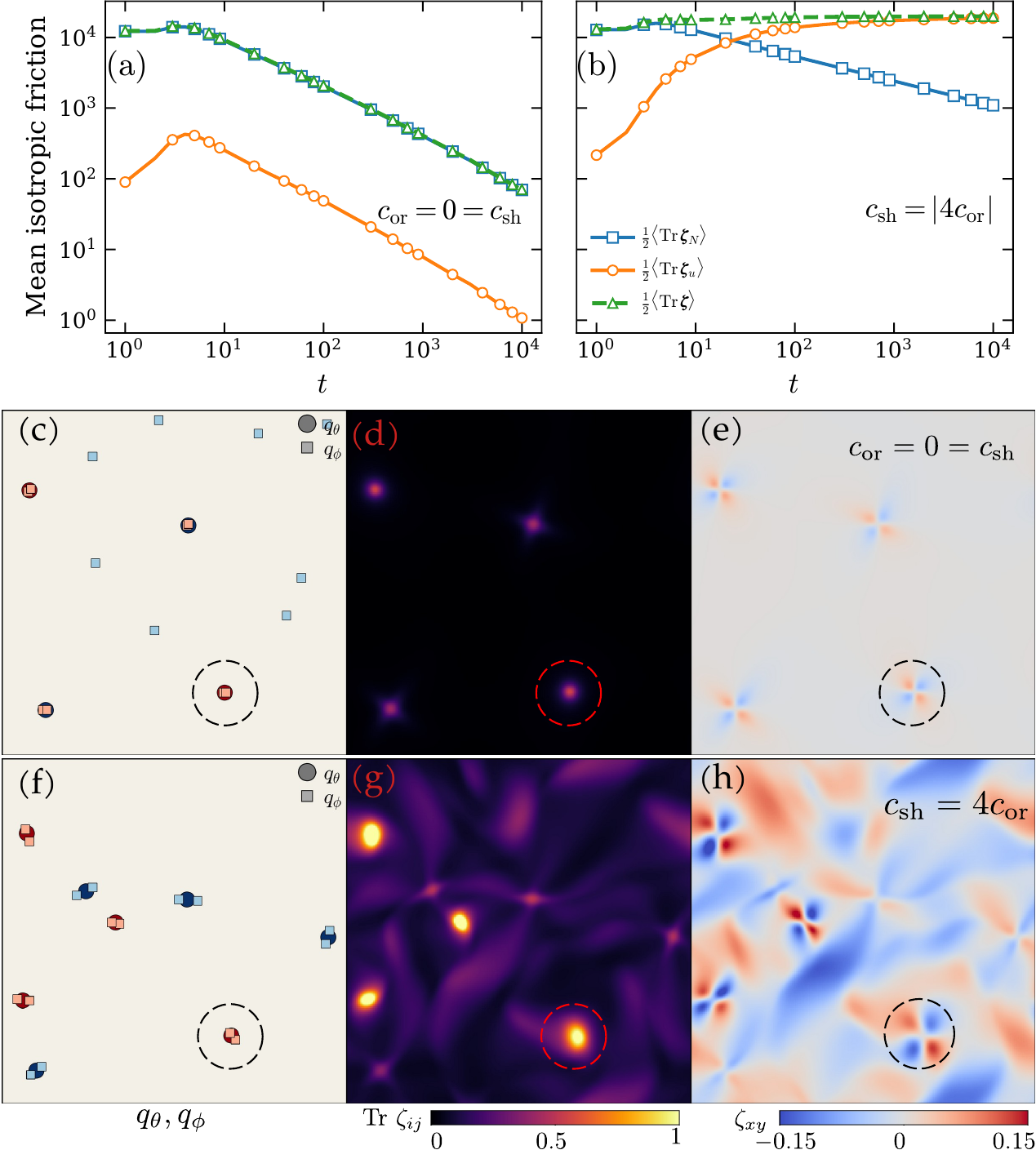}
    \caption{Dissipative mechanics of magnetoelasticity.
(a,b) Mean isotropic friction,
$\frac{1}{2}\mathrm{Tr}\,\boldsymbol{\zeta}$ (triangles),
resolved into corresponding magnetic ($\boldsymbol{\zeta}_N$, squares) and elastic ($\boldsymbol{\zeta}_u$, circles) contributions [Eqs.~\eqref{dissipative-power}-\eqref{zeta_Nu}]. For isotropic coupling $c_{\rm sh}=0=c_{\rm or}$ (a), dissipation is dominated by magnetic defect cores. In the presence of orientational coupling $c_{\rm sh}=2=4 c_{\rm or}$ (b), elastic friction generates much larger dissipation at sufficiently late times. Parameters: $L_x=512=L_y, dx=1, dt=0.01, c_{\rm iso}=1$. Other parameters same as in Fig.~\ref{fig:defects_all}. Data averaged over $40$ initial realizations.
(c--h) Representative defect configurations (c,f),
$\mathrm{Tr}\,\boldsymbol{\zeta}$ (d,g), and off-diagonal component
$\zeta_{xy}$ (e,h), for the corresponding two regimes at $t=200$~\cite{Supp_Mov_2}. The quadrupolarly coupled, charge-locked regime exhibits a stronger and spatially extended elastic frictional response. Dashed circles highlight representative bound-defect complexes. Parameters for (c-h) same as in Fig.~\ref{fig:defects_all}. Color bar of (c,f) same as in Fig.~\ref{fig:defects_all}(c,f,i).
}\label{fig:friction}
\vspace{-0.2cm}
\end{figure}

Our main result is that these two different forms of defect bindings are accompanied by distinct phase-ordering kinetics. First, isotropic magnetostriction locally binds magnetic and elastic defects, while leaving magnetic coarsening essentially XY-like. In sharp contrast, quadrupolar magnetoelastic coupling produces charge-locked composite defects and, in the regimes considered in this work, an {\em ultraslow} coarsening of both magnetic and elastic sectors. Our numerical simulations of the relaxational dynamics [Eq.~\eqref{eq:dyn}] show that at late times both defect densities decay as $t^{-2/5}$, while the mean defect velocities decay as $t^{-4/5}$, consistent with a dynamic exponent $z\simeq 5$ [Fig.~\ref{fig:scaling}]. Remarkably, the same slow kinetics is observed whether the bound elastic charges have the same or the opposite sign as the magnetic defect. Thus defect binding alone does not determine ordering kinetics. To gain physical insight into the anomalous relaxation, we examine the dissipative response of the coupled fields [Fig.~\ref{fig:friction}]. While isotropic magnetostriction exhibits a predominantly magnetic, core-localized frictional response, quadrupolar coupling generates a strong, spatially extended elastic contribution that accompanies the ultraslow defect dynamics.

\medskip

{\it Continuum theory of magnetoelasticity:}
We consider a two-dimensional magnetoelastic medium described by an in-plane magnetic order parameter ${\bf N}({\bf r},t)=(N_x,N_y)$ and an elastic displacement field ${\bf u}({\bf r},t)=(u_x,u_y)$.  We assume purely relaxational dynamics for the two vector fields,
\begin{align}\label{eq:dyn}
\partial_t {\bf N} = -\Gamma_{N} \frac{\delta F}{\delta {\bf N}}, \quad
\partial_t {\bf u} = -\Gamma_{u} \frac{\delta F}{\delta {\bf u}},
\end{align}
where $\Gamma_{N}$ and $\Gamma_u$ are mobilities, and the free energy $F$ couples magnetic order to deformations of the medium.  

For the free energy, we write~\cite{RevModPhys.21.541, Callen1963, Callen1965}
\begin{equation}\label{F_total}
F[{\bf N},{\bf u}]=\int d^2{\bf r} \Big( f_{\rm el}[{\bf u}]+f_{\rm mag}[{\bf N}]+f_{\rm c}[{\bf N},{\bf u}] \Big),
\end{equation}
where $f_{\rm el}=\mu\,u_{ij}u_{ij}+\frac{\lambda}{2}\,u_{ii}u_{ii}$ and $f_{\rm mag} = \frac{a}{2}N^2+\frac{b}{4}N^4+ \frac{v}{6} N^6 + \frac{k_1}{2}(\nabla\cdot{\bf N})^2+\frac{k_2}{2}(\partial_i N_j)^2$ are the bare elastic and magnetic free energy densities. The magnetoelastic coupling is encoded in the contribution
\begin{align}
f_{\rm c} \equiv&\, \Big[f_{\rm c}^{\rm iso}\Big] + \Big[f_{\rm c}^{\rm quad}\Big] =  \Big[ c_{\rm iso}\, N^2 \, (u_{xx}+u_{yy}) \Big] \label{Fc} \\
&+ \Big[ c_{\rm or} \, (N_x^2-N_y^2) \, (u_{xx}-u_{yy}) + c_{\rm sh} \, N_xN_y \, u_{xy} \Big], \nonumber
\end{align}
where $c_{\rm iso}, c_{\rm or}, c_{\rm sh}$ are coupling constants, and $f_{\rm c}^{\rm iso}$ and $f_{\rm c}^{\rm quad}$ are the isotropic and quadrupolar parts of the magnetoelastic coupling, respectively, that we will discuss in detail later. We have $u_{ij} =\frac{1}{2}\left[\partial_i u_j + \partial_j u_i \right]$ and $N^2 = {\bf N} \cdot {\bf N}$. 

The elastic free energy is the harmonic theory of an isotropic solid, with Lam\'e coefficients $\mu$ and $\lambda$ controlling shear and compression. The magnetic contribution is a soft-spin XY-like theory: for $a<0$ and $b>0$ the local potential favors a state of finite magnetic order, while the gradient terms penalize spatial variations of its magnitude and orientation. We assume $v>0$ to ensure thermodynamic stability for cases when magnetoelastic coupling can change the sign of the coefficient of the quartic term in $N$. In the absence of magnetoelastic coupling, relaxation of this sector therefore proceeds through the familiar annihilation of magnetic vortices~\cite{Mazenko1997}. In the rest of the paper, we choose $a=-1=-b, k_1=0.5=k_2, v=1, \Gamma_N=1=\Gamma_u, \mu=1=2\lambda$ such that our model parameters are the magnetoelastic couplings $c_{\rm iso}, c_{\rm or}$ and $c_{\rm sh}$. We initialize $N_\alpha$ and $u_\alpha$ independently from uniform distributions on $[-1,1]$ and $[-10^{-2},10^{-2}]$, respectively.

\medskip

{\em Charge-locking:} In this model, the central physics is contained in Eq.~\eqref{Fc}, which describes the magnetoelastic coupling. The isotropic term $c_{\rm iso}$ couples the magnetic-order magnitude to the local dilation $u_{xx}+u_{yy}$, and represents isotropic magnetostriction; being orientation-independent, it cannot distinguish vortices from anti-vortices. The remaining terms, $c_{\rm or}$ and $c_{\rm sh}$, couple the magnetic quadrupole to shear strain and are sensitive to the relative magnetic-elastic orientation, thereby allowing topological-charge locking. These orthorhombic and shear couplings are not independent under continuous rotational symmetry. This becomes apparent when we introduce the magnetic and elastic quadrupolar fields
\begin{equation}
\mathbf{Q}_{m} = \left(N_x^2-N_y^2,\,2N_xN_y\right), \quad
\mathbf{Q}_{e} = \left(u_{xx}-u_{yy},\,2u_{xy}\right). \label{Qm_Qe}
\end{equation}
Writing the magnetic order parameter as
\begin{equation}
{\bf N}=N(\cos\theta,\sin\theta), \label{theta_def}
\end{equation}
the magnetic quadrupole takes the form $\mathbf Q_m =N^2(\cos2\theta,\sin2\theta)$. Similarly, the principal axes of the local strain define an orientational field $\phi$ through $\mathbf Q_e = Q_e(\cos2\phi,\sin2\phi)$, where $Q_e=\sqrt{(u_{xx}-u_{yy})^2+4u_{xy}^2}$. Thus, $\theta$ denotes the local orientation of the magnetic order, whereas
\begin{equation}\label{strain_orient}
\phi = \frac{1}{2} \tan^{-1} \left( \frac{2u_{xy}}{u_{xx}-u_{yy}} \right)
\end{equation}
specifies the orientation of a principal axis of the deviatoric strain. Since a principal axis is an unoriented line, it is invariant under $\phi\to\phi+\pi$, hence $\phi$ is a director angle. These two orientational fields support distinct topological defects. The magnetic angle $\theta$ is $2\pi$ periodic, so that its defects carry integer winding $q_\theta = \frac{1}{2\pi}\oint_{\mathcal C} d\theta = \pm 1$. By contrast, the identification $\phi\equiv\phi+\pi$ allows the elastic director to wind by $\pi$ around a closed contour, giving elementary half-integer defects, $q_\phi = \frac{1}{2\pi}\oint_{\mathcal C} d\phi = \pm\frac{1}{2}$. The magnetoelastic couplings determine how these two species of topological defects are related.

In terms of the orientation variables $\theta$ and $\phi$ defined in Eqs.~\eqref{Qm_Qe}-\eqref{strain_orient}, the quadrupolar magnetoelastic coupling in Eq.~\eqref{Fc} can be written as
\begin{align}
f_{\rm c}^{\rm quad} = \frac{N^2Q_e}{2} \Bigg[& \left(c_{\rm or}+\frac{c_{\rm sh}}{4}\right) \cos 2(\theta-\phi) \nonumber\\
& + \left(c_{\rm or}-\frac{c_{\rm sh}}{4}\right) \cos 2(\theta+\phi) \Bigg]. \label{fc_aniso}
\end{align}
This form makes transparent the distinct roles of isotropic and orientational magnetoelastic couplings in Eq.~\eqref{Fc}. First we note that the isotropic term $f_{\rm c}^{\rm iso}$ in Eq.~\eqref{Fc} depends only on the magnitude $N^2$ of the magnetic order and is therefore insensitive to the magnetic orientation $\theta$. Figures~\ref{fig:defects_all}(a-c) show that, in the absence of orientational couplings, both magnetic vortices and antivortices bind to pairs of $+1/2$ elastic defects~\cite{Supp_Mov}. Technical details on defect identification and tracking are provided in~\cite{PhysRevE.85.011153, Skogvoll2023, banerjee2026defectstatisticspulsatingactive} and Sec.~S1 of~\cite{Supp}; a mathematical justification of this charge-independent binding is given in Sec.~S2. The $-1/2$ elastic defects are not bound to the magnetic defects. Annihilation of a magnetic vortex–antivortex pair leaves a net elastic charge $+1$, from four $+1/2$ and two $-1/2$ defects~\cite{Supp_Mov}.

The quadrupolar couplings, by contrast, directly couple the magnetic orientation to the principal-axis director of the strain. This can be seen by minimizing the quadrupolar free energy, which yields (Sec. S3 in ~\cite{Supp})
\begin{equation}
\tan 2\phi = \frac{c_{\rm sh}}{4c_{\rm or}}\tan 2\theta , \label{tanphi}
\end{equation}
so that, upon encircling a magnetic vortex, the winding of the elastic director is determined by the sign of $c_{\rm or}c_{\rm sh}$. 

In the rotationally invariant limit, $c_{\rm sh}=4c_{\rm or}$, the $\theta+\phi$ channel vanishes and $f_{\rm c}^{\rm quad}=c_{\rm or}N^2Q_e\cos 2(\theta-\phi)=c_{\rm or}\,\mathbf Q_m\cdot\mathbf Q_e $. The sign of $c_{\rm or}$ determines whether $\theta$ and $\phi$ prefer to align parallel or perpendicular, but in either case their relative angle is fixed, so that $ \oint_{\mathcal C} d\phi=\oint_{\mathcal C} d\theta$ and $q_\phi=q_\theta$. Since the elementary defects of the director field carry charge $q_\phi=\pm1/2$, an integer magnetic vortex is therefore \emph{locked with} two elastic half-defects whose combined charge has the same sign as the magnetic defect [Fig.~\ref{fig:defects_all}(d-f)]. When the underlying crystal breaks continuous rotational symmetry, the ratio $c_{\rm sh}=4c_{\rm or}$ need not hold, and the additional $\theta+\phi$ channel in Eq.~\eqref{fc_aniso} opens up. In the complementary limit $c_{\rm sh}=-4c_{\rm or}$, the $\theta-\phi$ channel vanishes and $f_{\rm c}^{\rm quad}=c_{\rm or}N^2Q_e\cos 2(\theta+\phi)$. Energy minimization then fixes $\theta+\phi$ to a constant, giving $\oint_{\mathcal C} d\phi=-\oint_{\mathcal C} d\theta$ and $q_\phi=-q_\theta$. The elastic dressing therefore carries a total charge opposite to that of the magnetic defect [Fig.~\ref{fig:defects_all}(g-i)]. Thus, crystalline anisotropy can do more than modify the preferred relative orientation of magnetization and strain, i.e., it can reverse the topological charge locking between the elastic and magnetic defects. Therefore, the study of magnetic and elastic defects, together with their topological locking, can provide direct insights into the underlying free-energy landscape of a magnetoelastic material. In contrast to the topologically charged annihilation of the isotropically coupled magnetoelastic situation, anisotropic magnetoelastic coupling leads to a topologically neutral annihilation of a vortex-anti-vortex pair involving a pair each of $\pm 1/2$ elastic defects. In this paper, we focus our discussions on $4c_{\rm or}/c_{\rm sh}=\pm1$. The general anisotropic theory of topology of magnetoelasticity will be presented in a forthcoming article~\cite{Banerjee_Jaiswal}.

\medskip

{\it Ultraslow coarsening of magnetic domains:}
We now investigate the phase-ordering kinetics of the magnetoelastic coupling regimes characterized above, following a quench to zero temperature. We start from disordered magnetic and elastic initial states and simulate their evolution following Eq.~\eqref{eq:dyn}. The ensuing phase ordering proceeds through the motion and annihilation of magnetic vortices and elastic half-defects, allowing us to follow directly how magnetoelastic coupling modifies the coarsening kinetics. 

In the absence of quadrupolar couplings, $c_{\rm or}=0=c_{\rm sh}$, with $c_{\rm iso}\ne0$, the magnetic defects recover the well-known coarsening law of the two-dimensional XY model. Given that the total number of defects at an instant is $N_\theta(t)$, the average defect density $\langle n_\theta(t) \rangle = \frac{\langle N_\theta (t) \rangle}{L_xL_y}$ scales as  $\frac{\ln t}{t}$~\cite{PhysRevE.47.1525}, despite the binding of magnetic and elastic defects [black circles in Fig.~\ref{fig:scaling}(a)]. The behavior changes strikingly once the quadrupolar coupling is switched on. In the two charge-locked regimes, $4c_{\rm or}/c_{\rm sh}=\pm1$, the magnetic and elastic defect populations approach a common algebraic decay,
\begin{equation}
\langle n_\theta(t) \rangle, \, \langle n_\phi(t) \rangle \sim t^{-2/5},
\label{eq:anomalous_density}
\end{equation}
over the late-time coarsening regime [squares and triangles in Fig.~\ref{fig:scaling} (a)]. Interpreting the mean defect separation as $\ell(t)\sim \langle n_\theta\rangle^{-1/2}(t)$, gives $\ell(t)\sim t^{1/5}$, corresponding to a dynamic exponent
\begin{equation}
z \simeq 5.
\end{equation}
Therefore, charge-selective defect locking induced by orientational magnetoelastic coupling is associated with a substantially slower growth of magnetic orientational order. The inset in Fig.~\ref{fig:scaling}(a) reveals that the elastic defect density $\langle n_\phi \rangle$ follows the same temporal scaling as the magnetic one for both isotropic and anisotropic magnetoelastic coupling. We find that the appearance of the ultraslow coarsening regime coincides with the condition $\langle n_\phi \rangle \simeq 2\langle n_\theta \rangle$.

A complementary signature of slow kinetics is provided by the magnetic defect motion itself. We measure the spatially averaged speeds of magnetic and elastic defects during relaxation~\cite{Supp}.  Assuming coarsening is governed by a single growing length $\ell(t)\sim t^{1/z}$, the characteristic defect speed should scale as $\dot{\ell}(t) \sim t^{1/z-1}$. For $z=5$, this predicts a $t^{-4/5}$ decay for the defect speed. In fact, the defect-density and speed measurements can be combined without assuming a particular value of $z$. Using the spatially averaged defect speed  $\langle v_\theta \rangle$ as a proxy for the characteristic defect-velocity magnitude,
single-length coarsening implies $\langle v_\theta \rangle \sim\ell/t$ with $\ell\sim n_\theta^{-1/2}$, and therefore
\begin{equation}
\langle v_\theta\rangle\sqrt{\langle n_\theta \rangle} \sim 1/t.
\label{eq:velocity_density_collapse}
\end{equation}
Indeed, we observe that rescaling the magnetic-defect velocities according to Eq.~\eqref{eq:velocity_density_collapse} collapses the data across different magnetoelastic couplings at sufficiently late times [Fig.~\ref{fig:scaling}(b)]. Thus, independently measured defect numbers and velocities consistently bear signatures of a common slowly growing length scale in the presence of orientational magnetoelastic couplings. Importantly, we obtain $z \simeq 5$ for both $c_{\rm sh} = \pm 4 c_{\rm or}$. Thus, the same anomalous dynamic exponent is observed in both charge-locking regimes, despite the reversal of the elastic charges bound to magnetic vortices and antivortices. 
Another testable signature of this ultraslow coarsening is given by $\langle n_\theta \rangle \sim \sqrt{\langle v_\theta \rangle}$ [squares, triangles in Fig.~\ref{fig:scaling}(c)]. Hence, the information contained in Figs.~\ref{fig:scaling}(b)-(c) provides numerical evidence for the anomalous dynamic exponent $z\simeq5$.

\medskip

{\it Dissipative mechanics of slow coarsening:} While an analytical derivation of $z=5$ remains nontrivial, to understand the physical origin of the ultraslow kinetics, we examine how orientational magnetoelastic interactions modify the dissipative response of the coupled fields.
For relaxational dynamics, assuming rigid translation of a field texture with velocity $\mathbf{\bar v}$ gives the dissipated power~\cite{PhysRevLett.30.230, 10.1039/d1sm01335a} (also Sec. S4 in \cite{Supp})
\begin{equation}\label{dissipative-power}
\mathcal{P}_{\rm diss}= \bar {\rm v}_i \zeta_{ij} \bar {\rm v}_j , \qquad \zeta_{ij} = \zeta^{N}_{ij}+\zeta^{u}_{ij}
\end{equation}
with
\begin{equation}\label{zeta_Nu}
\zeta^{N}_{ij} = \frac{1}{\Gamma_N}\!\int\! d^2r\, \partial_i\mathbf{N} \cdot \partial_j\mathbf{N}, \quad
\zeta^{u}_{ij} = \frac{1}{\Gamma_u}\!\int\! d^2r\, \partial_i\mathbf{u} \cdot \partial_j\mathbf{u}.
\end{equation}
We characterize the direction-averaged response by the isotropic friction $\zeta_{\rm iso} = \frac12\mathrm{Tr}\,\boldsymbol{\zeta}$. As shown in Fig.~\ref{fig:friction}(a), when $c_{\rm or}=c_{\rm sh}=0$ the dissipation is predominantly magnetic (blue squares), while the elastic contribution remains small (orange circles). The total contribution (green triangles) decays strongly during coarsening.
We observe a qualitative change in this behavior on the charge-locked line
$c_{\rm sh}=|4c_{\rm or}|$ [Fig.~\ref{fig:friction}(b)]. While the magnetic friction still decays with time, the elastic contribution grows strongly, and overtakes its magnetic counterpart. The total isotropic friction remains large throughout the coarsening regime. 

The snapshots in Fig.~\ref{fig:friction}(c--h) reveal the corresponding reorganization of dissipation: without orientational locking, friction is concentrated primarily near isolated vortex cores [Fig.~\ref{fig:friction}(c-e)], whereas the topologically locked state develops a strong, spatially extended dissipative texture [Fig.~\ref{fig:friction}(f-h)]. The plot of the off-diagonal component $\zeta_{xy}$ reveals that this response is intrinsically tensorial rather than a simple enhancement of a scalar drag. In the charge locked regime, the magnetic defects have to move through a complex friction landscape that eventually slows down their annihilation. The ultraslow coarsening is, therefore, a signature of the anisotropic redistribution of dissipation into an extended elastic sector. In Appendix~\ref{app:adiabatic}, we show how orientational couplings necessarily lead to effective long-range interactions between magnetic spins.


\medskip

{\it Discussion:} Our results on inter-field defect binding and locked defect kinetics strongly suggest that defect structure and kinetics may be used as an inverse probe of otherwise difficult-to-access free energy landscapes [Fig.~\ref{fig:defects_all}]. Moreover, we reveal that charge selective locking between topological defects appears with a drastic slowdown of the magnetic coarsening process with an interesting dynamic exponent $z \simeq 5$. 

We carefully note that the growth law with $z \simeq 5$ has previously appeared in the literature, e.g., in stripe forming  systems~\cite{PhysRevLett.75.2152}, quenched two-dimensional Bose gases~\cite{Karl_2017}, vortex glasses in three-dimensions~\cite{Nattermann01072000} or in systems with higher order conservation laws~\cite{x8v2-rxhp}. While generally the physical origin of such $z\simeq 5$ growth law  is not well-understood~\cite{PhysRevLett.75.2152}, in some cases it has been assigned to vortex clustering~\cite{Karl_2017}. In contrast, we show that our $z \simeq 5$ scaling appears with charge-selective topological bindings of the two orientation fields and large anisotropic elastic friction. 

Closely related examples of coupling-induced slow coarsening are provided by Uchida~\cite{PhysRevE.66.040902} and Vats {\it et al}~\cite{Vats_2019}. Uchida found that orientational coarsening in fluid membranes is limited by the subdiffusive relaxation of membrane curvature, while Vats \textit{et al.} demonstrated slaved coarsening in ferronematics arising from magneto-nematic coupling. In our system, the ultraslow coarsening regime is accompanied by a pronounced dissipation in the elastic sector, highlighting the importance of the coupled magnetic and elastic dynamics. 

Our work thus belongs to a broader class of systems in which coupling distinct orientational or topological sectors reorganizes defect dynamics. Some recent examples include ferronematic systems~\cite{ferronematics-2025}, active solids~\cite{xv94-xpz2}, and studies of defect braiding and binding in active nematics~\cite{PhysRevLett.132.228301, jqkw-69d1}. Our results demonstrate that binding alone does not determine ordering kinetics. In the quadrupolarly coupled regimes considered here, ultraslow coarsening is accompanied by charge-selective topological locking with $\langle n_\phi \rangle \simeq 2\langle n_\theta \rangle$ [Fig.~\ref{fig:scaling}], and a pronounced elastic contribution to the dissipative response [Fig.~\ref{fig:friction}]. 

Verifying our predictions requires simultaneous resolution of the strain texture and magnetic configuration. Recent developments in electron microscopy allow for such nanoscale mapping of local strain and magnetization~\cite{Kang2025}. Tracking magnetic defects with a reconstruction of the elastic principal axes will allow for investigation of the local magnetoelastic textures. The kinetic signatures can be obtained by direct tracking of defects. Other interesting experimental avenues include controlling the motion of magnetic vortices by time-dependent strain~\cite{PhysRevLett.133.146701},  defect-mediated reorientation of magnetic textures driven by uniaxial stresses in thin films~\cite{PhysRevLett.134.136704}, or manipulation of intra-layer magnetic ordering through inter-layer shear couplings~\cite{Zhou2022}.

Within our minimal approach, we have ignored the role of temperature. We expect temperature could weaken defect binding, and thus make the dynamic exponent deviate from $z \simeq 5$. Further, we have restricted ourselves to situations where $c_{\rm or}^2 = \frac{c_{\rm sh}^2}{16}$. In a forthcoming  work~\cite{Banerjee_Jaiswal}, we will show in detail how the dynamic exponent $z$ and the charge-selective locking between topological defects are directly controlled by $c_{\rm or}$ and $c_{\rm sh}$. 

Our findings can potentially motivate new routes to controlling topological defects~\cite{doi:10.1073/pnas.2400933121} and exploring symmetry-dependent defect interactions in exciting altermagnetic systems~\cite{halder2026entropydrivenaltermagnetismthermalmagnons}. Another intriguing direction is to investigate analogous composite defects and their dynamics in  magnetized~\cite{Mertelj2013, doi:10.1126/science.aaw8719, PhysRevE.92.062133} and spin-polarized fluids~\cite{Florkowski:2017ruc, Bhadury:2022ulr}.


\medskip

\begin{acknowledgements}
We gratefully acknowledge stimulating discussions with Anamitra Mukherjee. We thank the Department of Atomic Energy (DAE), India, for financial support.
\end{acknowledgements}

\makeatletter
\let\normal@hangfrom@section\@hangfrom@section
\let\normal@sectioncntformat\@sectioncntformat
\makeatother

\appendix
\setcounter{secnumdepth}{2}
\section{Adiabatic elimination of elastic degrees of freedom}
\label{app:adiabatic}

We now briefly show how orientational magnetoelastic couplings $c_{\rm or}$ and $c_{\rm sh}$ lead to effective long-ranged interactions between the magnetic degrees of freedom. A detailed derivation is presented in Sec. S5 of \cite{Supp}. It is useful to examine the effective magnetic free energy obtained when the elastic degrees of freedom relax rapidly compared with the magnetic texture. In this adiabatic limit the displacement field is at mechanical equilibrium for a given magnetic configuration, $\delta F/\delta u_i=0$.

We introduce
\begin{equation}
S=N^2,\qquad Q_1=N_x^2-N_y^2,\qquad Q_2=2N_xN_y ,
\end{equation}
in terms of which the magnetostrictive stress is (Sec. S5 in \cite{Supp})
\begin{equation}
\Sigma_{ij} =
\begin{pmatrix}
c_{\rm iso}S+c_{\rm or}Q_1 & \dfrac{c_{\rm sh}}{4}Q_2\\[4pt]
\dfrac{c_{\rm sh}}{4}Q_2 & c_{\rm iso}S-c_{\rm or}Q_1
\end{pmatrix}.
\label{eq:stress_app}
\end{equation}
For a nonzero wave vector ${\bf q}$, mechanical equilibrium gives
\begin{equation}
D_{ij}({\bf q})u_j({\bf q})=J_i({\bf q}), \qquad J_i=-iq_j\Sigma_{ij},
\end{equation}
where
\begin{equation}
D_{ij} = \mu q^2 P^{T}_{ij} + (\lambda+2\mu)q^2P^{L}_{ij}.
\end{equation}
with $P^L_{ij}=q_iq_j/q^2$ and $P^T_{ij}=\delta_{ij}-P^L_{ij}$ denoting the longitudinal and transverse projection operators, respectively.

Eliminating the Gaussian displacement field~\cite{PhysRevLett.122.218002, PhysRevE.99.053004}, therefore, gives
\begin{equation}
F_{\rm eff}[{\bf N}] = F_{\rm mag}[{\bf N}] - \frac{1}{2} \int_{\bf q} J_i(-{\bf q}) D^{-1}_{ij}({\bf q}) J_j({\bf q}), \label{eq:Feff_compact}
\end{equation}
with
\begin{equation}
D^{-1}_{ij} = \frac{P^T_{ij}}{\mu q^2} + \frac{P^L_{ij}}{(\lambda+2\mu)q^2}.
\end{equation}

The physical distinction between isotropic and orientational magnetostriction becomes particularly transparent by defining
\begin{equation}
\Phi_c({\bf q}) = \frac{K_c({\bf q})}{q^2},
\end{equation}
where
\begin{equation}
K_c = c_{\rm or}(\partial_x^2-\partial_y^2)Q_1 + \frac{c_{\rm sh}}{2}\partial_x\partial_y Q_2 . \label{eq:Kc_app}
\end{equation}

Here, $K_c=\partial_i\partial_j\Sigma_{ij}^{\rm aniso}$ is the double divergence of the anisotropic magnetostrictive stress, while $\Phi_c$ is the corresponding nonlocal field, satisfying $\nabla^2\Phi_c=-K_c$.

One then obtains (Sec. S5 of \cite{Supp})
\begin{align}
F_{\rm eff} ={}&F_{\rm mag} \!- \frac{c_{\rm iso}^2}{2(\lambda+2\mu)} \!\int\! d^2r\,S^2 \!+ \frac{\lambda+\mu}{2\mu(\lambda+2\mu)} \!\int_{\bf q}|\Phi_c({\bf q})|^2 \nonumber\\
&-\frac{1}{2\mu} \int d^2r \left( c_{\rm or}^2Q_1^2 +\frac{c_{\rm sh}^2}{16}Q_2^2 \right) \nonumber\\
&+ \frac{c_{\rm iso}}{\lambda+2\mu} \int_{\bf q}S(-{\bf q})\Phi_c({\bf q}) . \label{eq:Feff_app}
\end{align}
Equation~\eqref{eq:Feff_app} highlights the qualitative difference between the couplings considered in the main text. For purely isotropic magnetostriction, $c_{\rm or}=c_{\rm sh}=0$, one has $K_c=0$, and integrating out elasticity only renormalizes the local quartic magnetic free energy,
\begin{equation}
F_{\rm eff}^{\rm iso} = F_{\rm mag} - \frac{c_{\rm iso}^2}{2(\lambda+2\mu)} \int d^2r\,N^4,
\end{equation}
and no nonlocal orientational interaction is generated.

By contrast, for the full charge-locking cases $c_{\rm sh}=\pm4c_{\rm or}$, the local anisotropic contribution reduces to an isotropic term since $Q_1^2+Q_2^2=S^2$. However, only $c_{\rm sh}=4c_{\rm or}$ preserves continuous rotational symmetry, whereas $c_{\rm sh}=-4c_{\rm or}$ retains crystalline anisotropy. In both cases, the nonlocal interactions survive, since $K_c$ is generally nonzero and takes the form
\begin{equation}
K_c^{\pm}\equiv K_c\big|_{c_{\rm sh}=\pm4c_{\rm or}}
=c_{\rm or}\left[
(\partial_x^2-\partial_y^2)Q_1
\pm2\partial_x\partial_yQ_2
\right].
\label{eq:Kpm_app}
\end{equation}
This provides a microscopic distinction between charge-blind isotropic magnetostriction and the orientational couplings associated with the ultraslow defect kinetics reported in the main text.


\bibliography{refs-top}


\clearpage
\onecolumngrid

\makeatletter
\let\@hangfrom@section\normal@hangfrom@section
\let\@sectioncntformat\normal@sectioncntformat
\@removefromreset{equation}{section}
\makeatother

\setcounter{secnumdepth}{2}

\setcounter{section}{0}
\setcounter{subsection}{0}
\setcounter{equation}{0}
\setcounter{figure}{0}
\setcounter{table}{0}
\setcounter{page}{1}

\renewcommand{\thesection}{S\arabic{section}}
\renewcommand{\thesubsection}{\thesection.\arabic{subsection}}

\renewcommand{\theequation}{S\arabic{equation}}
\renewcommand{\thefigure}{S\arabic{figure}}
\renewcommand{\thetable}{S\arabic{table}}

\begin{center}
\underline{\bfseries\Large Supplemental Material}\\[0.5cm]

{\bfseries\large
Charge-selective topological locking and ultraslow coarsening in magnetoelastic media
}\\[0.3cm]

Tirthankar Banerjee and Amaresh Jaiswal\\[0.2cm]

{\itshape
School of Physical Sciences, National Institute of Science Education and Research, An OCC of Homi Bhabha National Institute, Jatni-752050, India
}
\end{center}


\section{Topological charge densities and defect velocities}

We characterize defects in the magnetic and elastic orientational fields through the angles
\begin{equation}
\theta = \tan^{-1}\!\left(\frac{N_y}{N_x}\right), \qquad
\phi = \frac{1}{2}\tan^{-1}\!\left(\frac{2u_{xy}}{u_{xx}-u_{yy}}\right),
\end{equation}
where $\theta$ is the orientation of the magnetic order parameter, while $\phi$ denotes the principal axis of the deviatoric strain. Since $\phi$ is a director, $\phi\equiv\phi+\pi$, and it is convenient to introduce the single-valued angle
\begin{equation}
\alpha = 2\phi .
\end{equation}

For a generic angular field $X$, we define the corresponding two-component unit field
\begin{equation}
{\bf p}_X = (\cos X,\sin X).
\end{equation}
We can therefore associate a topological charge density with ${\bf p}_X$ as
\begin{equation}
\rho_X = \frac{1}{\pi} \left[ (\partial_x\cos X)(\partial_y\sin X) - (\partial_y\cos X)(\partial_x\sin X) \right]. \label{eq:rho_topological}
\end{equation}
Accordingly,
\begin{equation}
\rho_\theta = \rho_X\big|_{X=\theta},
\end{equation}
whereas for the elastic director
\begin{equation}
\rho_\phi = \frac{1}{2}\rho_\alpha, \qquad \alpha=2\phi . \label{eq:rho_phi}
\end{equation}
The factor $1/2$ accounts for the half-integer topological charge of elementary defects in the director field.

The local defect velocity is obtained from the continuity of the topological charge density. For a generic angular field $X$, the two components of the velocity field are evaluated as~\cite{banerjee2026defectstatisticspulsatingactive, PhysRevE.85.011153, Skogvoll2023}
\begin{align}
v_x^{(X)} &= \frac{ (\partial_t\sin X)(\partial_y\cos X) - (\partial_t\cos X)(\partial_y\sin X) } {\pi\rho_X}, \label{eq:vx_topological}\\
v_y^{(X)} &= \frac{ (\partial_t\cos X)(\partial_x\sin X) - (\partial_t\sin X)(\partial_x\cos X) } {\pi\rho_X}. \label{eq:vy_topological}
\end{align}
The corresponding local speed is
\begin{equation}
v_X = \sqrt{ \left(v_x^{(X)}\right)^2 + \left(v_y^{(X)}\right)^2 }.
\end{equation}
For magnetic defects we set $X=\theta$. For elastic defects we apply Eqs.~\eqref{eq:vx_topological} and \eqref{eq:vy_topological} to $X=\alpha=2\phi$. Multiplying both the topological density and current by $1/2$ leaves their ratio, and hence the velocity, unchanged.

Spatial derivatives are evaluated using centered finite differences. The temporal derivatives are evaluated from two configurations separated by one integration step,
\begin{equation}
\partial_t f(t) \simeq \frac{f(t)-f(t- d t)}{d t}.
\end{equation}
To restrict the velocity measurement to defect cores, we retain only lattice sites satisfying
\begin{equation}
|\rho_X|(dx)^2 \geq \rho_{\rm cut},
\end{equation}
where $dx$ is the spatial grid spacing with $\rho_{\rm cut}=0.08$ for both the magnetic and elastic sectors. The reported mean defect speed $\langle v_X\rangle$ is the arithmetic mean of the local speed over all sites satisfying this criterion.

We emphasize that the total number of defects is determined independently from the winding of the angular field around elementary lattice plaquettes. For each plaquette, the angular differences along its four directed edges are first mapped to the principal interval $(-\pi,\pi]$ in order to remove artificial jumps associated with the $2\pi$ periodicity of the angle. The topological charge is then obtained from the total accumulated phase change around the plaquette,
\begin{equation}
q_X = \frac{1}{2\pi} \sum_{\square} \Delta X ,
\end{equation}
where the sum runs counterclockwise over the four plaquette edges. A nonzero value of $q_X$ identifies a defect core within that plaquette.

For the magnetic angle $X=\theta$, this procedure gives integer charges $q_\theta=\pm1$. For the elastic director we instead apply the same construction to $\alpha=2\phi$, and define
\begin{equation}
q_\phi=\frac{1}{2}q_\alpha ,
\end{equation}
so that the elementary elastic defects carry half-integer charge $q_\phi=\pm1/2$. The total defect numbers $N_\theta$ and $N_\phi$ are obtained by counting plaquettes with nonzero $q_\theta$ and $q_\phi$, respectively. To eliminate numerical round-off errors, we round the computed winding numbers $q_\theta$ and $q_\alpha$ to the nearest integers, before obtaining the elastic charge as $q_\phi=q_\alpha/2$.

\subsection{Numerical integration scheme}

We integrate the coupled relaxational equations [Eq. (1) of the main text] for the magnetic order parameter $\mathbf N=(N_x,N_y)$ and the elastic displacement field $\mathbf u=(u_x,u_y)$ on a two-dimensional square lattice with spacing $dx$ and periodic boundary conditions.

For a generic field $f_{i,j}$, the first-order spatial derivatives are evaluated using second-order central finite differences,
\begin{align}
(\partial_x f)_{i,j} &\equiv \frac{f_{i+1,j}-f_{i-1,j}}{2~d x}, \\
(\partial_y f)_{i,j} &\equiv \frac{f_{i,j+1}-f_{i,j-1}}{2 ~d x}.
\end{align}
The Laplacian is discretized using the standard five-point stencil,
\begin{equation}
(\nabla^2 f)_{i,j} \equiv \frac{ f_{i+1,j}+f_{i-1,j} +f_{i,j+1}+f_{i,j-1} -4f_{i,j} }{(dx)^2}.
\end{equation}
Mixed derivatives, as well as the gradient-of-divergence terms appearing in the equations of motion, are evaluated by successive applications of the central-difference operators. For example,
\begin{equation}
(\partial_x\partial_y f)_{i,j} \equiv \frac{ f_{i+1,j+1}-f_{i+1,j-1} -f_{i-1,j+1}+f_{i-1,j-1} }{4(dx)^2}.
\end{equation}
Similarly, for a vector field $\mathbf A=(A_x,A_y)$,
\begin{equation}
\nabla\cdot\mathbf A = \partial_x A_x+\partial_y A_y,
\end{equation}
and the terms $\partial_i(\nabla\cdot\mathbf A)$ are obtained by applying the  corresponding central-difference operator to the discretized divergence. Time integration is performed using an explicit forward-Euler scheme.


\section{Defect binding for isotropic magnetostriction}

We now consider the purely isotropic magnetostrictive coupling,
\begin{equation}
F_{c} = c_{\rm iso} \int d^2r\, N^2\,u_{ii}, \qquad c_{\rm iso}>0, \qquad c_{\rm or}=0=c_{\rm sh}.
\end{equation}
In contrast to the quadrupolar coupling proportional to $c_{\rm or}$, this term contains only the magnitude $N^2$ of the magnetic order parameter and is therefore insensitive to the local magnetic orientation $\theta$. In particular, it cannot distinguish between a magnetic vortex and an antivortex,
\begin{equation}
q_\theta=+1 \qquad\text{and}\qquad q_\theta=-1,
\end{equation}
provided that their core profiles $N^2(r)$ are otherwise identical.

Varying the free energy with respect to the displacement field gives a magnetostrictive contribution to the elastic force density proportional to
\begin{equation}
\mathbf f_{c_{\rm iso}} \propto \nabla N^2.
\end{equation}
For an approximately axisymmetric magnetic defect, the magnitude of the magnetic order parameter depends predominantly on the radial coordinate,
\begin{equation}
N^2=N^2(r),
\end{equation}
and hence
\begin{equation}
\nabla N^2 = \frac{dN^2}{dr}\,\hat{\mathbf r}.
\end{equation}
Thus, both a magnetic vortex and an antivortex act as the same radial source for the elastic displacement field. The resulting displacement near an isolated magnetic core can therefore be written, to leading order, as
\begin{equation}
\mathbf u(\mathbf r) = u(r)\,\hat{\mathbf r}.
\end{equation}

Let $\varphi$ denote the polar angle in real space. For such a radial displacement field, the anisotropic strain components are
\begin{align}
u_{xx}-u_{yy} &= \left( \frac{du}{dr}-\frac{u}{r} \right)\cos 2\varphi, \\
2u_{xy} &= \left( \frac{du}{dr}-\frac{u}{r} \right)\sin 2\varphi.
\end{align}
The elastic quadrupolar order parameter therefore takes the form
\begin{equation}
\mathbf Q_e = A(r) \left( \cos2\varphi, \sin2\varphi \right),
\end{equation}
where
\begin{equation}
A(r) = \frac{du}{dr}-\frac{u}{r}.
\end{equation}
If $A(r)>0$, the elastic director is locally
\begin{equation}
\phi=\varphi,
\end{equation}
whereas for $A(r)<0$ one obtains
\begin{equation}
\phi=\varphi+\frac{\pi}{2}.
\end{equation}
The sign of $A(r)$ therefore changes only the local orientation of the quadrupole by $\pi/2$ and does not change its winding. In either case, encircling the magnetic core once gives
\begin{equation}
\oint_{\mathcal C} d\phi = 2\pi,
\end{equation}
and hence
\begin{equation}
q_\phi^{\rm tot} = \frac{1}{2\pi} \oint_{\mathcal C} d\phi = +1.
\end{equation}
Importantly, this result is independent of the sign of the magnetic topological charge. Thus, within the radial-core approximation,
\begin{equation}
q_\theta=+1 \quad\Longrightarrow\quad q_\phi^{\rm tot}=+1,
\end{equation}
and
\begin{equation}
q_\theta=-1 \quad\Longrightarrow\quad q_\phi^{\rm tot}=+1.
\end{equation}
This is qualitatively different from the $c_{\rm or}$-dominated regime, where the relative orientation of the magnetic and elastic quadrupoles is directly locked and the sign of the elastic winding follows the sign of the magnetic winding.

Since the elementary defects of the elastic quadrupolar field carry half-integer charge,
\begin{equation}
q_\phi=\pm\frac12,
\end{equation}
the induced charge-$+1$ elastic texture can lower its defect self-energy by fractionalizing into two elementary positive half-defects,
\begin{equation}
q_\phi^{\rm tot}=+1 \longrightarrow +\frac12+\frac12.
\end{equation}
Consequently, the defect structures observed around the magnetic cores take the form
\begin{equation}
q_\theta=+1 \quad\longrightarrow\quad \left(+\frac12,+\frac12\right)_\phi,
\end{equation}
and, remarkably,
\begin{equation}
q_\theta=-1 \quad\longrightarrow\quad \left(+\frac12,+\frac12\right)_\phi.
\end{equation}
The two positive half-defects repel each other, but remain localized in the vicinity of the magnetic core because together they constitute the charge-$+1$ quadrupolar strain texture generated by the magnetostrictive source. Their finite separation can therefore be understood as arising from a competition between the mutual repulsion of the two $+1/2$ defects and their localization by the strain texture associated with the magnetic core.

The negative elastic half-defects have a different status. Since the $c_{\rm iso}$ coupling generates the same positive quadrupolar winding around both signs of magnetic defects, there is no corresponding local mechanism that binds a pair of $-1/2$ defects to a magnetic core. Nevertheless, for periodic boundary conditions the total topological charge of the elastic quadrupolar field must remain neutral,
\begin{equation}
\sum_j q_{\phi,j}=0.
\end{equation}
The positive half-defects localized around magnetic cores must therefore be accompanied by compensating negative half-defects elsewhere in the system. These $-1/2$ defects need not occur as bound pairs and can remain spatially separated from the magnetic defects.

This distinction becomes particularly apparent during the annihilation of a magnetic vortex--antivortex pair. Before annihilation, the two magnetic defects carry zero net magnetic charge,
\begin{equation}
(+1)_\theta+(-1)_\theta=0,
\end{equation}
but each magnetic core can carry its own positive elastic quadrupolar texture,
\begin{equation}
(+1)_\theta \left( +\frac12,+\frac12 \right)_\phi, \qquad
(-1)_\theta \left( +\frac12,+\frac12 \right)_\phi.
\end{equation}
Thus, the magnetic annihilation event is not locally neutral in the elastic sector. When the magnetic vortex and antivortex annihilate, the magnetostrictive cores that localized the positive elastic half-defects also disappear. Negative half-defects from the surrounding elastic defect population can subsequently approach the annihilation region and annihilate with positive half-defects according to
\begin{equation}
+\frac12-\frac12\longrightarrow0.
\end{equation}
Any remaining positive half-defects are no longer confined by the magnetic cores and, being like-signed, repel and move apart.

The $c_{\rm iso}$-dominated regime therefore represents a form of \emph{core-induced elastic fractionalization}, rather than a direct topological locking of magnetic and elastic charges. The isotropic magnetostrictive coupling associates the presence of a magnetic core with a positive elastic quadrupolar winding, but does not encode the sign of the magnetic vortex. As a result, magnetic vortex--antivortex annihilation can leave behind a transient population of elastic half-defects, providing a direct dynamical distinction between isotropic magnetostriction and the charge-selective defect locking produced by the quadrupolar $c_{\rm or}$ coupling.


\section{Anisotropic magnetoelasticity: Charge-locked defect composites}

Here we derive the relation between the winding of the magnetic orientation and that of the elastic director for arbitrary values of the quadrupolar magnetoelastic couplings $c_{\rm or}$ and $c_{\rm sh}$.

We write the in-plane magnetic order parameter as
\begin{equation}
{\bf N} = N(\cos\theta,\sin\theta),
\end{equation}
so that
\begin{equation}
N_x^2-N_y^2 = N^2\cos 2\theta, \qquad
N_xN_y = \frac{N^2}{2}\sin 2\theta.
\end{equation}
Similarly, we parameterize the deviatoric strain as
\begin{equation}
u_{xx}-u_{yy} = Q_e\cos 2\phi, \qquad
2u_{xy} = Q_e\sin 2\phi,
\end{equation}
where
\begin{equation}
Q_e = \sqrt{(u_{xx}-u_{yy})^2+4u_{xy}^2}.
\end{equation}
The angle $\phi$ specifies a principal axis of the local strain and is therefore a director angle, $\phi\equiv\phi+\pi$.

The quadrupolar part of the magnetoelastic coupling density,
\begin{equation}
f_{\rm c}^{\rm quad} = c_{\rm or}(N_x^2-N_y^2)(u_{xx}-u_{yy}) + c_{\rm sh}N_xN_yu_{xy},
\end{equation}
can consequently be written as
\begin{equation}
f_{\rm c}^{\rm quad} = N^2Q_e \left[ c_{\rm or}\cos2\theta\cos2\phi + \frac{c_{\rm sh}}{4}\sin2\theta\sin2\phi \right]. \label{eq:fcquad_general_SM}
\end{equation}
Using elementary trigonometric identities, Eq.~\eqref{eq:fcquad_general_SM} is equivalently
\begin{align}
f_{\rm c}^{\rm quad} = \frac{N^2Q_e}{2} \Bigg[ \left(c_{\rm or}+\frac{c_{\rm sh}}{4}\right) \cos2(\theta-\phi) + \left(c_{\rm or}-\frac{c_{\rm sh}}{4}\right) \cos2(\theta+\phi) \Bigg]. \label{eq:fcquad_sumdiff_SM}
\end{align}

For a given magnetic orientation $\theta$, the preferred elastic orientation can be obtained by minimizing Eq.~\eqref{eq:fcquad_general_SM} with respect to $\phi$. It is useful to define
\begin{equation}
A(\theta) = c_{\rm or}\cos2\theta, \qquad
B(\theta) = \frac{c_{\rm sh}}{4}\sin2\theta,
\end{equation}
such that
\begin{equation}
f_{\rm c}^{\rm quad} = N^2Q_e \left[ A(\theta)\cos2\phi + B(\theta)\sin2\phi \right]. \label{eq:fc_AB}
\end{equation}
Introducing
\begin{equation}
R(\theta) = \sqrt{A^2(\theta)+B^2(\theta)} = \sqrt{ c_{\rm or}^2\cos^22\theta + \frac{c_{\rm sh}^2}{16}\sin^22\theta },
\end{equation}
and an angle $\delta(\theta)$ defined through
\begin{equation}
A=R\cos\delta, \qquad B=R\sin\delta,
\end{equation}
Eq.~\eqref{eq:fc_AB} becomes
\begin{equation}
f_{\rm c}^{\rm quad} = N^2Q_eR(\theta) \cos\!\left[2\phi-\delta(\theta)\right],
\end{equation}
where
\begin{equation}
\delta(\theta) = \tan^{-1} \left( \frac{c_{\rm sh} \sin 2\theta}{4 c_{\rm or}\cos2\theta} \right). \label{eq:delta_def}
\end{equation}

For $R(\theta)>0$, minimization therefore requires
\begin{equation}
2\phi-\delta(\theta) = \pi \qquad (\mathrm{mod}\;2\pi),
\end{equation}
or
\begin{equation}
2\phi = \delta(\theta)+\pi \qquad (\mathrm{mod}\;2\pi). \label{eq:phi_delta}
\end{equation}
Equivalently, differentiation of Eq.~\eqref{eq:fcquad_general_SM} gives for the free energy minimum
\begin{equation}
-c_{\rm or}\cos2\theta\sin2\phi + \frac{c_{\rm sh}}{4}\sin2\theta\cos2\phi = 0,
\end{equation}
and hence
\begin{equation}
\tan2\phi = \frac{c_{\rm sh}}{4c_{\rm or}}\tan2\theta, \label{eq:tan_phi_theta}
\end{equation}
with the appropriate branch selected by Eq.~\eqref{eq:phi_delta}. The above equation corresponds to Eq.~\eqref{tanphi} of the main text.

The topological content of this relation follows from the winding of $\delta(\theta)$.  The two components entering Eq.~\eqref{eq:delta_def} define the vector
\begin{equation}
{\bf h}(\theta) = \left( c_{\rm or}\cos2\theta,\, \frac{c_{\rm sh}}{4}\sin2\theta \right). \label{eq:h_vector}
\end{equation}
As $\theta$ varies, ${\bf h}(\theta)$ traces an ellipse,
\begin{equation}
\frac{h_x^2}{c_{\rm or}^2} + \frac{h_y^2}{(c_{\rm sh}/4)^2} = 1,
\end{equation}
centered at the origin.  The angle $\delta$ is precisely the polar angle of this auxiliary vector.

The sense in which this ellipse is traversed can be seen directly by differentiating Eq.~\eqref{eq:delta_def}. Using
\begin{equation}
\frac{d\delta}{d\theta} = \frac{ h_x\,\partial_\theta h_y - h_y\,\partial_\theta h_x }{ h_x^2+h_y^2 },
\end{equation}
we obtain
\begin{equation}
\frac{d\delta}{d\theta} = \frac{ c_{\rm or}c_{\rm sh}/2 }{ c_{\rm or}^2\cos^22\theta + (c_{\rm sh}^2/16)\sin^22\theta }. \label{eq:delta_derivative}
\end{equation}
For $c_{\rm or}c_{\rm sh}\neq0$, and the denominator being positive, we have
\begin{equation}
\operatorname{sgn} \left( \frac{d\delta}{d\theta} \right) = \operatorname{sgn}(c_{\rm or}c_{\rm sh}).
\end{equation}
Thus the auxiliary ellipse winds in the same sense as the magnetic orientation when $c_{\rm or}c_{\rm sh}>0$, and in the opposite sense when $c_{\rm or}c_{\rm sh}<0$.

Consider now a magnetic defect of integer charge $q_\theta=m$, defined by
\begin{equation}
q_\theta = \frac{1}{2\pi}\oint_{\mathcal C} d\theta = m.
\end{equation}
Upon encircling the defect,
\begin{equation}
\Delta\theta = 2\pi m.
\end{equation}
Since ${\bf h}$ depends on $2\theta$, the ellipse Eq.~\eqref{eq:h_vector} is traversed $2|m|$ times. Its polar angle therefore changes by
\begin{equation}
\Delta\delta = 4\pi m\, \operatorname{sgn}(c_{\rm or}c_{\rm sh}). \label{eq:delta_winding}
\end{equation}
Using Eq.~\eqref{eq:phi_delta}, the corresponding change in the elastic director angle is
\begin{equation}
\Delta\phi = \frac{1}{2}\Delta\delta = 2\pi m\, \operatorname{sgn}(c_{\rm or}c_{\rm sh}).
\end{equation}
The elastic defect charge,
\begin{equation}
q_\phi = \frac{1}{2\pi}\oint_{\mathcal C} d\phi,
\end{equation}
therefore satisfies the general locking rule
\begin{equation}
q_\phi = \operatorname{sgn}(c_{\rm or}c_{\rm sh})\,q_\theta, \qquad c_{\rm or}c_{\rm sh}\neq0. \label{eq:general_charge_locking}
\end{equation}
Consequently,
\begin{equation}
c_{\rm or}c_{\rm sh}>0 \quad\Longrightarrow\quad q_\phi=q_\theta,
\end{equation}
whereas
\begin{equation}
c_{\rm or}c_{\rm sh}<0 \quad\Longrightarrow\quad q_\phi=-q_\theta.
\end{equation}

The continuously rotationally invariant case $c_{\rm sh}=4c_{\rm or}$ provides a simple check.  Equation~\eqref{eq:fcquad_sumdiff_SM} reduces to
\begin{equation}
f_{\rm c}^{\rm quad} = c_{\rm or}N^2Q_e\cos2(\theta-\phi),
\end{equation}
so that the minimum fixes
\begin{equation}
\phi = \theta + \mathrm{const.},
\end{equation}
and hence the magnetic and elastic windings have the same sign. The constant offset is $0$ or $\pi/2$ modulo $\pi$, depending on the sign of $c_{\rm or}$, but it does not affect the winding.

Similarly, for $c_{\rm sh}=-4c_{\rm or}$,
\begin{equation}
f_{\rm c}^{\rm quad} = c_{\rm or}N^2Q_e\cos2(\theta+\phi),
\end{equation}
and the minimum requires
\begin{equation}
\phi = -\theta+\mathrm{const.},
\end{equation}
giving oppositely signed magnetic and elastic windings.

Finally, the axes $c_{\rm or}=0$ and $c_{\rm sh}=0$ are singular boundaries between these two locking sectors.  When both coefficients are nonzero,
\begin{equation}
R(\theta) = \sqrt{ c_{\rm or}^2\cos^22\theta + \frac{c_{\rm sh}^2}{16}\sin^22\theta } >0
\end{equation}
for every $\theta$, so the preferred elastic orientation is well-defined continuously around a magnetic defect.  If $c_{\rm sh}=0$, however,
\begin{equation}
R(\theta) = |c_{\rm or}\cos2\theta|,
\end{equation}
which vanishes whenever $\cos2\theta=0$.  Likewise, for $c_{\rm or}=0$,
\begin{equation}
R(\theta) = \frac{|c_{\rm sh}|}{4}|\sin2\theta|,
\end{equation}
which vanishes whenever $\sin2\theta=0$.  At these orientations the quadrupolar coupling does not select a unique elastic director, and the map from $\theta$ to $\phi$ becomes ill-defined. The lines $c_{\rm or}c_{\rm sh}=0$ therefore separate two topologically distinct locking sectors:
\begin{equation}
\begin{array}{lll}
c_{\rm or}c_{\rm sh}>0 &:& \text{same-sign charge locking},\\[2mm]
c_{\rm or}c_{\rm sh}<0 &:& \text{opposite-sign charge locking},\\[2mm]
c_{\rm or}c_{\rm sh}=0 &:& \text{singular locking boundary}.
\end{array}
\end{equation}
%


\section{Dissipative cost of translating magnetoelastic textures}

The interpretation of $\zeta^{(N)}_{ij}$ and $\zeta^{(u)}_{ij}$ as friction tensors in Eq. (13) of the main text follows directly from the relaxational dynamics. For the magnetic field,
\begin{equation}
\partial_t N_\alpha = -\Gamma_N \frac{\delta F}{\delta N_\alpha},
\end{equation}
so that the corresponding dissipation rate is
\begin{equation}
-\dot F_N = \frac{1}{\Gamma_N} \int d^2r\, \sum_{\alpha=x,y} (\partial_t N_\alpha)^2.
\end{equation}
If the texture undergoes a rigid translation with velocity $\mathbf{v}$,
\begin{equation}
N_\alpha(\mathbf r,t) = N_\alpha^{(0)}(\mathbf r-\mathbf vt),
\end{equation}
then
\begin{equation}
\partial_t N_\alpha = -v_i\partial_i N_\alpha.
\end{equation}
Substitution into the dissipation rate gives
\begin{equation}
-\dot F_N = v_i\,\zeta^{(N)}_{ij}\,v_j,
\end{equation}
with
\begin{equation}
\zeta^{(N)}_{ij} = \frac{1}{\Gamma_N} \int d^2r\, \sum_{\alpha=x,y} (\partial_i N_\alpha)(\partial_j N_\alpha).
\end{equation}

The same argument for the displacement field,
\begin{equation}
\partial_t u_\alpha = -\Gamma_u \frac{\delta F}{\delta u_\alpha},
\end{equation}
yields
\begin{equation}
-\dot F_u = v_i\,\zeta^{(u)}_{ij}\,v_j,
\end{equation}
where
\begin{equation}
\zeta^{(u)}_{ij} = \frac{1}{\Gamma_u} \int d^2r\, \sum_{\alpha=x,y} (\partial_i u_\alpha)(\partial_j u_\alpha).
\end{equation}
Thus, $\zeta^{(N)}_{ij}$ and $\zeta^{(u)}_{ij}$ [Eq.~\eqref{zeta_Nu} of main text] quantify the dissipative cost of translating the magnetic and elastic textures, respectively, and therefore play the role of translational friction tensors.

\section{Adiabatic elimination of the elastic displacement field:
effective long-ranged magnetic interactions}
\label{sec:integrate_u_general}

We now integrate out the elastic displacement field ${\bf u}$ for the general magnetoelastic coupling. The total free energy is given by
\begin{equation}
F[{\bf N},{\bf u}] = F_{\rm mag}[{\bf N}] + F_{\rm el}[{\bf u}] + F_{\rm c}[{\bf N},{\bf u}], \label{F_total_general}
\end{equation}
where the elastic free energy is
\begin{equation}
F_{\rm el} = \int d^2r \left[ \mu u_{ij}u_{ij} + \frac{\lambda}{2}u_{ii}u_{ii} \right], \label{Fel_general}
\end{equation}
with
\begin{equation}
u_{ij} = \frac{1}{2} \left( \partial_i u_j+\partial_j u_i \right),
\end{equation}
and
\begin{equation}
F_{\rm c} = \int d^2r \left[ c_{\rm iso}~ N^2 (u_{xx}+u_{yy}) + c_{\rm or} ~ (N_x^2-N_y^2)(u_{xx}-u_{yy}) + c_{\rm sh} ~ N_xN_y u_{xy} \right]. \label{fc_general}
\end{equation}
%

\subsection{Magnetostrictive stress tensor}

It is convenient to define
\begin{equation}
S \equiv {\bf N}\cdot{\bf N} = N_x^2+N_y^2, \label{S_general}
\end{equation}
together with the two components
\begin{equation}
Q_1 \equiv N_x^2-N_y^2, \qquad
Q_2 \equiv 2N_xN_y. \label{Q12_general}
\end{equation}
These obey the identity
\begin{equation}
Q_1^2+Q_2^2=S^2. \label{Q_identity_general}
\end{equation}

We write the coupling in the form
\begin{equation}
F_{\rm c} = \int d^2r\, \Sigma_{ij}u_{ij}. \label{Fc_stress_general}
\end{equation}
Since
\begin{equation}
\Sigma_{ij}u_{ij} = \Sigma_{xx}u_{xx} + \Sigma_{yy}u_{yy} + 2\Sigma_{xy}u_{xy},
\end{equation}
comparison with Eq.~\eqref{fc_general} gives
\begin{equation}
\Sigma_{ij} =
\begin{pmatrix}
c_{\rm iso}S+c_{\rm or}Q_1
&
\dfrac{c_{\rm sh}}{4}Q_2
\\[3mm]
\dfrac{c_{\rm sh}}{4}Q_2
&
c_{\rm iso}S-c_{\rm or}Q_1
\end{pmatrix}.
\label{Sigma_general}
\end{equation}
Equivalently, it is useful to separate the isotropic and traceless parts,
\begin{equation}
\Sigma_{ij} = c_{\rm iso}S\,\delta_{ij} + R_{ij}, \label{Sigma_split}
\end{equation}
where
\begin{equation}
R_{ij} =
\begin{pmatrix}
c_{\rm or}Q_1
&
\dfrac{c_{\rm sh}}{4}Q_2
\\[3mm]
\dfrac{c_{\rm sh}}{4}Q_2
&
-c_{\rm or}Q_1
\end{pmatrix},
\qquad
R_{ii}=0.
\label{Rij_general}
\end{equation}
%

\subsection{Elastic theory in Fourier space}

We introduce the Fourier transform
\begin{equation}
u_i({\bf r}) = \int_{\bf q} u_i({\bf q}) e^{i{\bf q}\cdot{\bf r}}, \qquad
\int_{\bf q} \equiv \int\frac{d^2q}{(2\pi)^2}.
\end{equation}
For nonzero wave vectors,
\begin{equation}
u_{ij}({\bf q}) = \frac{i}{2} \left[ q_i u_j({\bf q}) + q_j u_i({\bf q}) \right].
\end{equation}
The elastic free energy becomes
\begin{equation}
F_{\rm el} = \frac{1}{2} \int_{\bf q} u_i(-{\bf q}) D_{ij}({\bf q}) u_j({\bf q}),
\end{equation}
where
\begin{equation}
D_{ij}({\bf q}) = \mu q^2\delta_{ij} + (\lambda+\mu)q_iq_j. \label{D_general}
\end{equation}

Introducing the longitudinal and transverse projectors,
\begin{equation}
P^L_{ij} = \frac{q_iq_j}{q^2}, \qquad
P^T_{ij} = \delta_{ij} - \frac{q_iq_j}{q^2},
\end{equation}
we may write
\begin{equation}
D_{ij} = (\lambda+2\mu)q^2P^L_{ij} + \mu q^2P^T_{ij},
\end{equation}
and hence
\begin{equation}
D^{-1}_{ij} = \frac{P^L_{ij}}{(\lambda+2\mu)q^2} + \frac{P^T_{ij}}{\mu q^2}. \label{Dinv_general}
\end{equation}

Using the symmetry of $\Sigma_{ij}$, the coupling can be written as
\begin{equation}
F_{\rm c} = -\int_{\bf q} u_i(-{\bf q})J_i({\bf q}),
\end{equation}
where
\begin{equation}
J_i({\bf q}) = -iq_j\Sigma_{ij}({\bf q}). \label{J_general}
\end{equation}
The part of the free energy depending on ${\bf u}$ therefore reads
\begin{equation}
F_u = \frac{1}{2} \int_{\bf q} u_i(-{\bf q})D_{ij}u_j({\bf q}) - \int_{\bf q} u_i(-{\bf q})J_i({\bf q}).
\end{equation}
Completing the square and integrating over the Gaussian displacement field gives
\begin{equation}
F_{\rm eff}[{\bf N}] = F_{\rm mag}[{\bf N}] - \frac{1}{2} \int_{\bf q} J_i(-{\bf q}) D^{-1}_{ij}({\bf q}) J_j({\bf q}). \label{Feff_general_master}
\end{equation}
%

\subsection{Longitudinal and transverse stress components}

Let
\begin{equation}
\hat{\bf q} = (\cos\psi,\sin\psi), \qquad \hat{\bf t} = (-\sin\psi,\cos\psi).
\end{equation}
The longitudinal projection of the stress is
\begin{equation}
\Sigma_{LL} \equiv \hat q_i\Sigma_{ij}\hat q_j.
\end{equation}
Using Eq.~\eqref{Sigma_general}, one obtains
\begin{equation}
\Sigma_{LL} = c_{\rm iso}S + c_{\rm or}Q_1\cos(2\psi) + \frac{c_{\rm sh}}{4} Q_2\sin(2\psi). \label{SigmaLL_general}
\end{equation}
Similarly, the transverse--longitudinal projection is
\begin{equation}
\Sigma_{TL} \equiv \hat t_i\Sigma_{ij}\hat q_j,
\end{equation}
which gives
\begin{equation}
\Sigma_{TL} = \frac{c_{\rm sh}}{4} Q_2\cos(2\psi) - c_{\rm or}Q_1\sin(2\psi). \label{SigmaTL_general}
\end{equation}
Since
\begin{equation}
J_i = -iq\left( \Sigma_{LL}\hat q_i + \Sigma_{TL}\hat t_i \right),
\end{equation}
Eq.~\eqref{Feff_general_master} becomes
\begin{align}
F_{\rm eff} = F_{\rm mag} - \frac{1}{2} \int_{\bf q} \Bigg( \frac{1}{\lambda+2\mu} \left| c_{\rm iso}S + c_{\rm or}Q_1\cos(2\psi) + \frac{c_{\rm sh}}{4}Q_2\sin(2\psi)
\right|^2 + \frac{1}{\mu} \left| \frac{c_{\rm sh}}{4}Q_2\cos(2\psi) - c_{\rm or}Q_1\sin(2\psi) \right|^2 \Bigg). \label{Feff_explicit_general}
\end{align}
Equation~\eqref{Feff_explicit_general} is the exact effective magnetic free energy obtained after integrating out the nonzero-wave-vector elastic displacement modes.

\subsection{Separation into local and nonlocal contributions}

To expose the nonlocal interaction, define
\begin{equation}
X({\bf q}) \equiv c_{\rm or}Q_1({\bf q}), \qquad
Y({\bf q}) \equiv \frac{c_{\rm sh}}{4}Q_2({\bf q}),
\end{equation}
and introduce
\begin{equation}
g({\bf q}) = X({\bf q})\cos(2\psi) + Y({\bf q})\sin(2\psi), \label{g_general}
\end{equation}
and
\begin{equation}
h({\bf q}) = Y({\bf q})\cos(2\psi) - X({\bf q})\sin(2\psi). \label{h_general}
\end{equation}
These satisfy
\begin{equation}
|g|^2+|h|^2 = |X|^2+|Y|^2. \label{gh_identity}
\end{equation}

We now define the generalized compatibility field in Fourier space by
\begin{equation}
\Phi_c({\bf q}) \equiv - \left[ c_{\rm or}Q_1({\bf q})\cos(2\psi) + \frac{c_{\rm sh}}{4} Q_2({\bf q})\sin(2\psi) \right]. \label{Phi_general}
\end{equation}
Thus
\begin{equation}
g({\bf q}) = -\Phi_c({\bf q}).
\end{equation}
Using Eq.~\eqref{gh_identity}, Eq.~\eqref{Feff_explicit_general} can be rearranged as
\begin{align}
F_{\rm eff} ={}& F_{\rm mag} - \frac{c_{\rm iso}^2}{2(\lambda+2\mu)} \int\! d^2r\,S^2 - \frac{1}{2\mu} \int\! d^2r \left[ c_{\rm or}^2Q_1^2 + \frac{c_{\rm sh}^2}{16}Q_2^2 \right] 
+ \frac{c_{\rm iso}}{\lambda+2\mu} \int_{\bf q}\! S(-{\bf q})\Phi_c({\bf q}) + \frac{\lambda+\mu}{2\mu(\lambda+2\mu)} \int_{\bf q}\! |\Phi_c({\bf q})|^2. \label{Feff_Phi_general}
\end{align}
This expression separates the effective free energy into local
magnetic terms and compatibility-mediated nonlocal interactions.

\subsection{Generalized incompatibility field}

The field $\Phi_c$ can be written in a coordinate-independent form. Since $R_{ii}=0$, Eq.~\eqref{Rij_general} gives
\begin{equation}
\Phi_c({\bf q}) = - \frac{q_iq_j}{q^2} R_{ij}({\bf q}). \label{Phi_R}
\end{equation}
We therefore define the generalized magnetic incompatibility
\begin{equation}
{\cal K}_c({\bf r}) \equiv \partial_i\partial_jR_{ij}({\bf r}). \label{Kc_definition}
\end{equation}
Its Fourier transform is
\begin{equation}
{\cal K}_c({\bf q}) = -q_iq_jR_{ij}({\bf q}),
\end{equation}
and hence
\begin{equation}
\Phi_c({\bf q}) = \frac{{\cal K}_c({\bf q})}{q^2}. \label{Phi_K_general}
\end{equation}

In two dimensions, Eq.~\eqref{Kc_definition} becomes
\begin{align}
{\cal K}_c ={}& c_{\rm or} (\partial_x^2-\partial_y^2) Q_1 + \frac{c_{\rm sh}}{2} \partial_x\partial_y Q_2 \nonumber\\
={}& c_{\rm or} (\partial_x^2-\partial_y^2) (N_x^2-N_y^2) + c_{\rm sh} \partial_x\partial_y (N_xN_y). \label{Kc_explicit}
\end{align}
Thus the relative values of $c_{\rm or}$ and $c_{\rm sh}$ determine the relative weight with which the normal-strain and shear sectors contribute to the elastic incompatibility generated by a magnetic texture.

\subsection{Real-space form of the effective interaction}

Introduce the Green function
\begin{equation}
V({\bf q}) = \frac{1}{q^2},
\end{equation}
which satisfies
\begin{equation}
-\nabla^2 V({\bf r}) = \delta^{(2)}({\bf r}),
\end{equation}
and the biharmonic Green function
\begin{equation}
V_2({\bf q}) = \frac{1}{q^4},
\end{equation}
satisfying
\begin{equation}
\nabla^4V_2({\bf r}) = \delta^{(2)}({\bf r}).
\end{equation}
In two dimensions,
\begin{equation}
V({\bf r}) = -\frac{1}{2\pi} \ln\left(\frac{r}{a_0}\right),
\end{equation}
while
\begin{equation}
V_2({\bf r}) = \frac{r^2}{8\pi} \ln\left(\frac{r}{a_0}\right),
\end{equation}
up to the usual additive or homogeneous contributions.

Equation~\eqref{Phi_K_general} implies
\begin{equation}
\Phi_c({\bf r}) = \int d^2r'\, V({\bf r}-{\bf r}') {\cal K}_c({\bf r}').
\end{equation}
Consequently, Eq.~\eqref{Feff_Phi_general} becomes
\begin{align}
F_{\rm eff} ={}& F_{\rm mag} - \frac{c_{\rm iso}^2} {2(\lambda+2\mu)} \int d^2r\,S^2({\bf r}) - \frac{1}{2\mu} \int d^2r \left[ c_{\rm or}^2 (N_x^2-N_y^2)^2 + \frac{c_{\rm sh}^2}{4} N_x^2N_y^2 \right] \nonumber\\
& + \frac{c_{\rm iso}}{\lambda+2\mu} \int d^2r\,d^2r'\, S({\bf r}) V({\bf r}-{\bf r}') {\cal K}_c({\bf r}') + \frac{\lambda+\mu}{2\mu(\lambda+2\mu)} \int d^2r\,d^2r'\, {\cal K}_c({\bf r}) V_2({\bf r}-{\bf r}') {\cal K}_c({\bf r}'). \label{Feff_real_general}
\end{align}
Equation~\eqref{Feff_real_general} is the effective magnetic free energy obtained after exactly integrating out the compatible nonzero-wave-vector elastic displacement modes. Using Eq.~\eqref{Kc_explicit}, we can write
\begin{align}
F_{\rm eff} ={}& F_{\rm mag} - \frac{c_{\rm iso}^2}{2(\lambda+2\mu)} \int d^2r\,S^2({\bf r}) - \frac{1}{2\mu} \int d^2r \left[ c_{\rm or}^2 (N_x^2-N_y^2)^2 + \frac{c_{\rm sh}^2}{4} N_x^2N_y^2 \right] \nonumber\\
& + \frac{c_{\rm iso} c_{\rm or}}{\lambda+2\mu} \int d^2r\,d^2r'\, S({\bf r}) V({\bf r}-{\bf r}') (\partial_x'^2-\partial_y'^2)(N_x^2({\bf r}') - N_y^2({\bf r}')) \nonumber\\
& + \frac{c_{\rm iso} c_{\rm sh}}{\lambda+2\mu} \int d^2r\,d^2r'\, S({\bf r}) V({\bf r}-{\bf r}') (\partial_{x'}\partial_{y'})(N_x({\bf r}')N_y({\bf r}')) \nonumber\\
& + \frac{\lambda+\mu}{2\mu(\lambda+2\mu)} \int d^2r\,d^2r'\, {\cal K}_c({\bf r}) V_2({\bf r}-{\bf r}') {\cal K}_c({\bf r}'). \label{Feff_real_general_2}
\end{align}

We now consider the special case $c_{\rm sh}=4c_{\rm or}$. For this choice, the local anisotropic contribution simplifies as
\begin{align}
c_{\rm or}^2(N_x^2-N_y^2)^2 + \frac{c_{\rm sh}^2}{4}N_x^2N_y^2 &= c_{\rm or}^2 \left[ (N_x^2-N_y^2)^2 + 4N_x^2N_y^2 \right] \nonumber\\
&= c_{\rm or}^2 (N_x^2+N_y^2)^2 = c_{\rm or}^2 S^2. \label{local_simplification}
\end{align}
Furthermore, the generalized incompatibility field becomes
\begin{align}
{\cal K}_c ={}& c_{\rm or} (\partial_x^2-\partial_y^2) (N_x^2-N_y^2) + 4c_{\rm or} \partial_x\partial_y(N_xN_y) \nonumber\\
={}& c_{\rm or}{\cal K}_N, \label{Kc_KN_relation}
\end{align}
where we have defined
\begin{equation}
{\cal K}_N = (\partial_x^2-\partial_y^2) (N_x^2-N_y^2) + 4\partial_x\partial_y(N_xN_y). \label{KN_definition}
\end{equation}
Substituting \eqref{Kc_KN_relation} into Eq.~\eqref{Feff_real_general}, we obtain
\begin{align}
F_{\rm eff} ={} F_{\rm mag} - \left[ \frac{c_{\rm iso}^2}{2(\lambda+2\mu)} + \frac{c_{\rm or}^2}{2\mu} \right] \int d^2r\,S^2({\bf r}) &+ \frac{c_{\rm iso}c_{\rm or}}{\lambda+2\mu} \int d^2r\,d^2r'\, S({\bf r}) V({\bf r}-{\bf r}') {\cal K}_N({\bf r}') \nonumber\\
&+ \frac{c_{\rm or}^2(\lambda+\mu)}{2\mu(\lambda+2\mu)} \int d^2r\,d^2r'\, {\cal K}_N({\bf r}) V_2({\bf r}-{\bf r}') {\cal K}_N({\bf r}'). \label{Feff_rotational}
\end{align}
Thus, on the line $c_{\rm sh}=4c_{\rm or}$, the local elasticity-induced orientational anisotropy vanishes. The anisotropic magnetoelastic coupling contributes only an isotropic local renormalization proportional to $S^2=N^4$, together with the compatibility-mediated nonlocal interaction governed by ${\cal K}_N$.

\subsection{Elasticity-induced orientational anisotropy}

An important difference from the rotationally invariant magnetoelastic coupling becomes apparent by writing
\begin{equation}
N_x=N\cos\theta, \qquad N_y=N\sin\theta.
\end{equation}
Then
\begin{equation}
Q_1=S\cos(2\theta), \qquad Q_2=S\sin(2\theta),
\end{equation}
and using \eqref{Feff_real_general}, the local contribution generated by the anisotropic elastic couplings becomes
\begin{align}
F_{\rm loc}^{({\rm an})} = -\frac{1}{2\mu} \int d^2r\, S^2 \left[ c_{\rm or}^2\cos^2(2\theta) + \frac{c_{\rm sh}^2}{16}\sin^2(2\theta) \right]. \label{Flocal_theta_1}
\end{align}
Using
\begin{equation}
\cos^2(2\theta) = \frac{1+\cos(4\theta)}{2}, \qquad
\sin^2(2\theta) = \frac{1-\cos(4\theta)}{2},
\end{equation}
this can be written as
\begin{align}
F_{\rm loc}^{({\rm an})} = -\frac{1}{4\mu} \int d^2r\,S^2 \Bigg[ c_{\rm or}^2 + \frac{c_{\rm sh}^2}{16} + \left( c_{\rm or}^2 - \frac{c_{\rm sh}^2}{16} \right) \cos(4\theta) \Bigg]. \label{Flocal_theta_2}
\end{align}
Therefore, independent values of $c_{\rm or}$ and $c_{\rm sh}$ generically generate a fourfold orientational anisotropy in the effective magnetic theory.  This {\em local} anisotropy disappears whenever
\begin{equation}
c_{\rm sh}^2 = 16c_{\rm or}^2.
\end{equation}
For $c_{\rm sh}=\pm4c_{\rm or}$, however, one obtains
\begin{equation}
{\cal K}_c = c_{\rm or} \left[ (\partial_x^2-\partial_y^2) (N_x^2-N_y^2) \pm 4\partial_x\partial_y(N_xN_y) \right]. \label{Kc_minus}
\end{equation}
Therefore, although both $c_{\rm sh}=4c_{\rm or}$ and $c_{\rm sh}=-4c_{\rm or}$ eliminate the local fourfold anisotropy, they produce different incompatibility fields and hence different nonlocal elastic interactions. Only the choice $c_{\rm sh}=4c_{\rm or}$ gives ${\cal K}_c\propto\partial_i\partial_jQ_{ij}$ and restores rotational invariance of the magnetoelastic system.


\end{document}